\documentstyle[12pt]{article}

\newcommand{\EQ}{\begin{equation}}
\newcommand{\EN}{\end{equation}}
\newcommand{\bea}{\begin{eqnarray}}
\newcommand{\ena}{\end{eqnarray}}

\renewcommand{\a}{\alpha}
\renewcommand{\b}{\beta}

\renewcommand{\d}{\delta}

\newcommand{\pa}{\partial}

\renewcommand{\L}{\Lambda}

\newcommand{\p}{\pi}

\renewcommand{\S}{\Sigma}

\begin{document}

\topmargin 0pt
\oddsidemargin 5mm

\renewcommand{\Im}{{\rm Im}\,}
\newcommand{\NP}[1]{Nucl.\ Phys.\ {\bf #1}}
\newcommand{\PL}[1]{Phys.\ Lett.\ {\bf #1}}
\newcommand{\NC}[1]{Nuovo Cimento {\bf #1}}
\newcommand{\CMP}[1]{Comm.\ Math.\ Phys.\ {\bf #1}}
\newcommand{\PR}[1]{Phys.\ Rev.\ {\bf #1}}
\newcommand{\PRL}[1]{Phys.\ Rev.\ Lett.\ {\bf #1}}
\newcommand{\MPL}[1]{Mod.\ Phys.\ Lett.\ {\bf #1}}
\renewcommand{\thefootnote}{\fnsymbol{footnote}}
\newpage
\begin{titlepage}
\vspace{2cm}
\begin{center}
{\bf{{\large QUANTUM BOUNDARY CURRENTS FOR }}} \\
{\bf{{\large NONSIMPLY--LACED TODA THEORIES}}} \\
\vspace{2cm}
{\large S. Penati, A. Refolli and D. Zanon} \\
\vspace{.2cm}
{\em Dipartimento di Fisica dell' Universit\`{a} di Milano and} \\
{\em INFN, Sezione di Milano, Via Celoria 16, I-20133 Milano, Italy}\\
\end{center}
\vspace{2cm}
\centerline{{\bf{Abstract}}}
\vspace{.5cm}
We study the quantum integrability of nonsimply--laced affine Toda theories
defined on the half--plane and explicitly construct the first nontrivial
higher--spin charges in specific examples. We find that, in contradistinction
to the classical case, addition of total derivative terms
to the "bulk" current plays a relevant role for the quantum boundary
conservation.

\vfill
\noindent
IFUM--518--FT \hfill {October 1995}
%\noindent
%\hep--th/9501105 \hfill
\end{titlepage}
\newpage
Two--dimensional quantum field theories defined on a manifold with boundary
are interesting for the description of various physical phenomena \cite{b1}.
If the boundary system is quantum integrable, an exact scattering matrix can
be constructed and the model is on--shell completely solvable \cite{b2}.
The existence of
an exact S matrix is guaranteed whenever the model possesses symmetries
generated by high--spin conserved charges.

{\it Classical} integrability has been studied for affine Toda theories based
on simply--laced as well as nonsimply--laced Lie algebras \cite{b3,corr}.
Recently we have addressed the issue of boundary conservation
at the {\it quantum} level \cite{b4}.
In particular we have considered the first relevant
quantum currents for the sinh--Gordon model and the $a^{(1)}_n$ Toda systems
defined on the half plane, perturbed by a boundary potential.
Here we extend the analysis to the case of nonsimply--laced affine Toda
theories. We have found that in order to ensure current conservation at the
quantum level, total derivative terms need to be added to the currents.
These terms, while irrelevant at the classical level, are crucial for the
construction of exact quantum symmetries of the theory. We explicitly
present the results for the spin--4 currents of the $d^{(2)}_3$ and $c^{(1)}_2$
theories.

\vspace{.4cm}

We work in euclidean space with the following notation for coordinates
\EQ
x = \frac{x_0 + ix_1}{\sqrt{2}} \qquad \quad \bar{x} =
\frac{x_0 - ix_1}{\sqrt{2}}
\EN
and derivatives
\EQ
\pa \equiv \pa_x = \frac{1}{\sqrt{2}} (\pa_0 -i\pa_1) \qquad ~~~
\bar{\pa} \equiv \pa_{\bar x} = \frac{1}{\sqrt{2}} (\pa_0 +i\pa_1) \qquad ~~~
\Box = 2 \pa \bar{\pa}
\EN

An affine Toda theory  based on a Lie algebra ${\cal G}$ of rank $N$,
has an exponential interaction of the form
\EQ
V = \sum_{j=0}^{N} q_j e^{\vec{\a}_j \cdot \vec{\phi}}
\label{1a}
\EN
where $\a_j$,
($j=1,\cdots,N$) are the simple roots of the Lie algebra and
$\a_0 = -\sum_{j=1}^N q_j \a_j$, with $q_j$ the Kac labels ($q_0 = 1$).
It has been shown that if one restricts the class of boundary potentials
\cite{b2,b3,corr} to
\EQ
B = \sum_{j=0}^N d_j e^{\frac12 \vec{\a}_j \cdot \vec{\phi}}
\label{1b}
\EN
with appropriate coefficients $d_j$, the
corresponding boundary Toda model defined by the action
\EQ
{\cal S} = \frac{1}{\b^2} \int d^2 x \left\{ \theta(x_1) \left[ \frac12
\pa_{\mu} \vec{\phi} \cdot \pa_{\mu} \vec{\phi} + V \right] -
\delta(x_1)B \right\}
\label{action}
\EN
is classically integrable. Indeed, in the upper half--plane $x_1 \ge 0$,
one can construct an infinite number of conserved charges as follows:
first one determines the currents which satisfy the classical conservation
laws in the "bulk" $x_1>0$
\EQ
\bar{\pa} J^{(n)} + \pa \Theta^{(n)} = 0  \qquad \quad
\pa \tilde{J}^{(n)} + \bar{\pa} \tilde{\Theta}^{(n)} = 0
\label{4}
\EN
Second one checks that the following boundary conditions are satisfied
\EQ
\left. J_1^{(n)} \right|_{x_1=0} \equiv
\left. i\left( J^{(n)} - \tilde{J}^{(n)} - \Theta^{(n)} + \tilde{\Theta}^{(n)}
\right) \right|_{x_1=0} = -\pa_0 \Sigma_0^{(n)}
\label{9}
\EN
with $\Sigma_0$ a local function of the fields evaluated at $x_1=0$.
Finally one obtains the corresponding conserved charges as
\EQ
q^{(n-1)} = \int_0^{+\infty} dx_1 J_0^{(n)} ~~+~ \Sigma_0^{(n)}
\label{8}
\EN
where $J_0^{(n)} = J^{(n)} + \tilde{J}^{(n)} + \Theta^{(n)} +
\tilde{\Theta}^{(n)}$.
Notice that the relations in (\ref{4}), (\ref{9}) are valid on--shell, i.e.
using the equations of motion in the bulk region
\EQ
\Box \vec{\phi} =  \sum_{j=0}^N q_j \vec{\a}_j e^{\vec{\a}_j \cdot
\vec{\phi}}
\label{2}
\EN
and at the boundary
\EQ
\left. \frac{\pa \phi_a}{\pa x_1}\right|_{x_1=0} = - \frac{\pa B}{\pa \phi_a}
\label{3}
\EN

As a final remark we observe that given $J^{(n)}$, $\Theta^{(n)}$,
$\tilde{J}^{(n)}$, $\tilde{\Theta}^{(n)}$
satisfying the conservation equations in (\ref{4}), equivalent sets
of currents can be constructed adding total derivative terms
\EQ
J^{(n)} \rightarrow J^{(n)}+\pa U~~~~~~~,~~~~~~~~
\Theta^{(n)}\rightarrow \Theta^{(n)} -\bar{\pa}U
\label{total}
\EN
Indeed (\ref{4}) hold unmodified, while
\bea
&& J^{(n)}_0 \rightarrow J^{(n)}_0-i \sqrt{2}~\pa_1(U-\tilde{U})~~~~~~~~~~~~~~
J^{(n)}_1 \rightarrow J^{(n)}_1+i \sqrt{2}~\pa_0(U-\tilde{U}) \nonumber\\
&~&~~~~~~~~~~~~~~~~~~~~~~~~~~~~~~~~~~~\nonumber \\
&& \Sigma_0^{(n)} \rightarrow  \left. \Sigma_0^{(n)}-i \sqrt{2}~
(U-\tilde{U}) \right|_{x_1=0}
\ena
Consequently the corresponding charges are conserved and identical
to the ones in (\ref{8}). Thus at the classical level total derivative
terms in the currents are undetermined and not relevant for the
conservation laws. This situation changes completely in the
quantum case to which we turn now.

In order to extend the above analysis to the quantum level it is convenient to
recast the problem in a perturbation theory approach, so that classical results
correspond to tree level calculations and quantum corrections are given by
loop contributions. In the following we use the same techniques
described in \cite{b5,b4}, which allow to obtain exact, all loop--order
results.
Here we only summarize the clue steps and the main formulas.

The classical conservation equations in (\ref{4}) and (\ref{9}) are
reexpressed in perturbation theory as
\EQ
\bar{\pa} \left\langle J^{(n)}(x,\bar{x}) \right\rangle \equiv \bar{\pa}
\left\langle J^{(n)}(x,\bar{x}) ~e^{- {\cal S}_{i}^V} \right\rangle_0 =
{}~-\pa \left\langle \Theta^{(n)} \right\rangle
\label{38}
\EN
for $x_1>0$ and
\EQ
\left. \left\langle J^{(n)}_1(x,\bar{x}) \right\rangle \right|_{x_1=0}
\equiv \left. \left\langle J^{(n)}_1(x,\bar{x}) ~e^{- {\cal S}_{i}}
\right\rangle_0 \right|_{x_1=0} = ~-\pa_0 \left\langle \S_0\right\rangle
\label{39}
\EN
at $x_1=0$. \\
We have defined the interaction term
${\cal S}_{i}\equiv {\cal S}_{i}^V+{\cal S}_{i}^B$, where
 ${\cal S}_{i}^V =  \frac{1}{\b^2}
\int_{-\infty}^{+\infty} dx_0 \int_0^{+\infty} dx_1 V$, with $V$ the
affine Toda potential (\ref{1a}) and ${\cal S}^B_{i} = -\frac{1}{\b^2}
\int_{-\infty}^{+\infty} dx_0 B$, with $B$ the boundary perturbation
(\ref{1b}).
The calculations are then performed using massless propagators
\EQ
G_{ij}(x,x') = -\frac{\b^2}{4\pi} \delta_{ij}
\left[ \log{2|x-x'|^2} + \log{2|x-\bar{x}'|}^2 \right]
\label{41}
\EN
and normal ordering the exponentials in $V$ and $B$
so that no ultraviolet divergences are produced.

A theory defined on a manifold with boundary is quantum integrable
if one can show that the conservation equations in
(\ref{38}), (\ref{39}) are not spoiled by anomalies.
Anomalous contributions would arise if Wick contractions
of the currents with the interaction exponentials produce
{\em local} terms which cannot be written as total derivatives.
Moreover local contributions arise only if the calculation
of the l.h.s. of eqs. (\ref{38}), (\ref{39}) produces enough
two--dimensional and one--dimensional $\d$--functions,
so that the integrations in ${\cal S}_{i}$ might be performed
explicitly.

Thus we proceed through a series of subsequent
steps: first we consider the conservation
equation in the bulk region, (\ref{38}).
At this stage the current $J^{(n)}$ is given by the most general expression
of spin $n$, i.e. a
sum of terms containing $n$ $\pa$--derivatives of the fields
with coefficients to be determined.
Then we compute to all--loop
orders the local contributions that arise from
$\bar{\pa} \left\langle J^{(n)} \right\rangle $.
It is easy to
realize that it is sufficient to expand the exponential
to first order in ${\cal S}_{i}^V$, since only one
two--dimensional $\d$--function can be produced in the course of
this calculation.
Indeed using in the half plane the relation
\EQ
\bar{\pa}_x \frac{1}{(x-w)^k} = \frac{2\pi}{(k-1)!} \pa_w^{k-1} \delta^{(2)}
(x-w)
\label{43}
\EN
we obtain local expressions. They are of two kinds:
total $\pa$--derivative contributions which give rise to the trace,
and terms not expressible as $\pa$--derivatives which
must vanish in order not to produce anomalies. The yet undetermined
coefficients in the current must be chosen so to cancel these
potentially anomalous terms. In this fashion
the quantum current $J^{(n)}$ and its corresponding quantum trace
$\Theta^{(n)}$ defined for $x_1>0$ can be computed exactly.

The second part of the calculation involves directly the boundary
perturbation. The quantum expressions just obtained
for $J^{(n)}$ and $\Theta^{(n)}$ in the bulk are used in
$J_1^{(n)}=i(J^{(n)}-\tilde{J}^{(n)}-\Theta^{(n)}+\tilde{\Theta}^{(n)})$
and then the condition (\ref{39}) is imposed at the boundary.
Again one is searching for potential anomalies, i.e.
local terms which are not $\pa_0$--derivatives. In this case in order
to isolate these local contributions one needs consider
higher--order terms in the expansion of the interaction potential,
at most first order in V, but possibly higher order in B. Indeed
repeated use of the relation
\EQ
\lim_{x_1 \to 0^+} \left( \frac{1}{(x_0-w_0 -ix_1)^k} -
\frac{1}{(x_0-w_0+ix_1)^k} \right) = \frac{2 \pi i}{(k-1)!} \pa_{w_0}^{k-1}
\delta(x_0 -w_0)
\label{47}
\EN
might allow to perform several integrations along the
one--dimensional boundary, thus producing local terms.

As a final step one has to analyze the local contributions which cannot
be written as $\pa_0$--derivatives
of suitable expressions, and understand whether they
correspond to real boundary anomalies.

This briefly describes the general procedure.
Now we want to examine
the role played by total derivative terms that, as we have anticipated,
become relevant at the quantum level.
The addition of a $\pa U$ term to the $J^{(n)}$ current
modifies the quantum conservation condition in the bulk by
a term $\bar{\pa} \langle \pa U \rangle=
\pa \bar{\pa}\langle U\rangle $. As explained
above one computes the local terms from $\bar{\pa}\langle U \rangle$
and identifies the corresponding contribution to the quantum trace.
Clearly no anomaly is produced, being the result
automatically in the form $\pa \L$, but while the tree level (classical)
contributions in $\L$ are equal to $\bar{\pa} U$, the loop
(quantum) corrections are not reexpressible in general as
$\bar{\pa}$--derivatives. Obviously this means that
these terms might lead to  quantum corrections in $J^{(n)}_1$ which
are not $\pa_0$--derivatives and therefore
affect the boundary condition (\ref{39}) in a nontrivial manner.

In this letter we present the quantum results for two cases,
namely the
spin--4 current conservation for the
$d_3^{(2)}\equiv a_3^{(2)}$ and the $c_2^{(1)}$ nonsimply--laced
Toda models. A general and more detailed discussion will be
reported elsewhere \cite{b6}.

\vspace{.8cm}
{\em{The $d^{(2)}_3$ affine Toda theory}}

\vspace{.2cm}
\noindent
This nonsimply--laced theory is described by the lagrangian
\EQ
\b^2 {\cal L}= \theta(x_1) \left[ \frac12  \pa_{\mu} \phi_1 \pa_{\mu} \phi_1
+\frac12  \pa_{\mu} \phi_2 \pa_{\mu} \phi_2  + e^{-\phi_1-\phi_2}
+  e^{2\phi_1} + e^{-\phi_1 +\phi_2}\right] -\d(x_1) B
\label{action1}
\EN
with $B$ given by
\EQ
B= d_0~ e^{- \frac12 (\phi_1+\phi_2 )} +d_1~  e^{\phi_1} +
d_2~ e^{-\frac12 (\phi_1 -\phi_2)}
\label{pot1}
\EN
The first nontrivial high--spin conserved current in the bulk region
is at spin 4, with a general form including total derivative terms
(based also on the symmetry of the lagrangian under
$\phi_2\rightarrow -\phi_2$)
\bea
J^{(4)}&=&A(\pa \phi_1)^2(\pa \phi_2)^2 +B(\pa \phi_1)^4+C(\pa \phi_2)^4
+D\pa\phi_1 \pa\phi_2\pa^2 \phi_2 +E(\pa^2 \phi_1)^2
+F(\pa^2 \phi_2)^2 + \nonumber\\
{}~&~& +G\pa((\pa \phi_2)^2 \pa \phi_1)
+H\pa(( \pa \phi_1)^3) + I\pa(\pa\phi_1\pa^2\phi_1)
+J\pa(\pa\phi_2\pa^2\phi_2)
\label{current1}
\ena
The quantum conservation of this current for the theory defined
on the whole two--dimensional plane has been studied in detail
in Ref. \cite{b5}, where the coefficients $A, B, \dots, F$ have been
determined (with $\a \equiv \frac{\b^2}{2\p}$)
\bea
A &=& 1+\frac{\a}{2}~~~~~~~~~~~~~~~~B=C=-\frac{\a}{12}~~~~~~~~~
{}~~~~~~~~~D=2+3\a+\a^2  \nonumber\\
E &=& -\frac{\a}{12}(1+3\a+\a^2)~~~~~~~~~~~~~~~~~
F=1+\frac{23}{12}\a+\a^2+\frac{\a^3}{6}
\label{sol1}
\ena
As emphasized above the conservation equation (\ref{38}) does
not impose any restriction on
the constants $G, H, I, J$. We have computed the quantum trace
and found
\EQ
\Theta =\Theta_0+\Theta_1+\Theta_2
\label{trace}
\EN
where
\bea
\Theta_0&=&\left\{ \left[ -\frac{\a}{8}(1+\frac{31}{18}\a+\frac{5}{6}\a^2
+\frac{1}{9}\a^3) + \frac{\a^2}{16} (G+H) - \frac{\a}{8} (I+J) \right]
(\pa\phi_1+\pa\phi_2)^2+ ~\right. \nonumber\\
{}~&~&~+\left[ \frac{\a}{4} G+
\frac{3}{2}(1+\frac{\a}{2})H -\frac{1}{2}I \right] (\pa\phi_1)^2~+ \nonumber\\
{}~&~&~+\left[ \frac{1}{2}(1+\a)G -\frac{1}{2} J\right]
(\pa\phi_2)^2+ \nonumber\\
{}~&~&~+\left[ (1+\frac{3}{4}\a)G+
\frac{3}{4}\a H -\frac{1}{2} (I+J)\right]
\pa\phi_1 \pa\phi_2+ \nonumber\\
{}~&~&~+\left[ -\frac{\a}{24}(1+\frac{5}{6}\a-\frac{\a^2}{2}
-\frac{\a^3}{3})-\frac{\a^2}{16}(G+H) +\frac{\a}{8} (I+J)\right]
(\pa^2\phi_1+\pa^2\phi_2 )
+\nonumber\\
{}~&~&~+\left. \frac{1}{2} I ~\pa^2\phi_1+ \frac{1}{2} J~ \pa^2\phi_2 \right\}
e^{-\phi_1-\phi_2}\nonumber\\
&~&~~~~~~~~~~\nonumber \\
\Theta_1 &=&\left\{ \left[ -\frac{\a}{6}(1+\frac{10}{3}\a+3\a^2
+\frac{2}{3} \a^3) -(3+6\a+2\a^2)H -2(1+\a)I\right] (\pa\phi_1)^2+\right.
\nonumber\\
{}~&~&~+\left[ (1+\frac{3}{2}\a+\frac12\a^2)-G \right] (\pa\phi_2)^2+
\nonumber\\
{}~&~&~+\left. \left[ -\frac{\a^2}{36}(1+3\a+2\a^2)-\a^2 H
-(1+\a)I\right]
\pa^2\phi_1 \right\} e^{2\phi_1}\nonumber\\
&~&~~~~~~~~~~~~~\nonumber \\
\Theta_2&=& \Theta_0(\phi_2\rightarrow -\phi_2)
\label{trace1}
\ena
The quantum expressions in (\ref{current1}), (\ref{sol1}) and (\ref{trace1})
are then used to compute $J^{(4)}_1$, and finally the boundary
condition (\ref{39}) is imposed. The local terms from
the computation of the l.h.s. of (\ref{39}) which are not total
$\pa_0$--derivatives of suitable expressions,
group themselves into two separate sets,
terms containing three $\pa_0$--derivatives and terms with one
$\pa_0$--derivative. The first set of terms automatically
adds up to zero once use is made of the specific form of the
boundary potential in (\ref{pot1}). Absence of anomalies requires also
the cancellation of the second set of terms and this leads to the following
equations
\bea
{}~&~&\frac{\a}{2} (G-3H)d_0=0 \nonumber\\
{}~&~&~~~~~~~~~~~~~~~~\nonumber \\
{}~&~&(3+2\a-\a^2)d_1^2 d_0 -\left[ 6+10\a+\frac{8}{3}\a^2-\frac{13}{9}\a^3
-\a^4 -\frac{2}{9}\a^5+ \right. \nonumber\\
{}~&~&~~~~~~~~~~~~~~~~~~~~~~~\left. +4\a(3+3\a+\a^2)H+4\a(1+\a)I\right]
d_0=0 \nonumber\\
{}~&~&~~~~~~~~~~~~~~~~\nonumber \\
{}~&~&\frac{\a}{4}d_0^2 d_1+\frac{\a}{2} \left[ 1+\frac{5}{2}\a +\frac{55}{36}
\a^2+\frac{5}{12}\a^3+\frac{\a^4}{18}+(1+\a+\frac{\a^2}{4})G+ \right.
\nonumber\\
{}~&~&~~~~~~~~~~~~~~~~~~~~~~\left.
+(3+3\a+\frac{\a^2}{4})H-(2+\frac{1}{2}\a)I-\frac{1}{2}\a J \right] d_1
=0 \nonumber\\
{}~&~&~~~~~~~~~~~~~~~~\nonumber \\
{}~&~&\frac{\a}{2}(G-3H)d_2=0 \nonumber\\
{}~&~&(3+2\a-\a^2)d_1^2d_2 - \left[ 6+10\a+\frac{8}{3}\a^2-\frac{13}{9}\a^3
-\a^4-\frac{2}{9}\a^5+ \right. \nonumber\\
{}~&~&~~~~~~~~~~~~~~~~~~~~~~\left. +4\a(3+3\a+\a^2)H+4\a(1+\a)I\right]
d_2=0 \nonumber\\
{}~&~&~~~~~~~~~~~~~~~~\nonumber \\
{}~&~&\frac{\a}{4}d_2^2 d_1+\frac{\a}{2} \left[ 1+\frac{5}{2}\a +\frac{55}{36}
\a^2+\frac{5}{12}\a^3+\frac{\a^4}{18}+(1+\a+\frac{\a^2}{4})G+ \right.
\nonumber\\
{}~&~&~~~~~~~~~~~~~~~~~~~~~~\left.
+(3+3\a+\frac{\a^2}{4})H-(2+\frac{1}{2}\a)I-\frac{1}{2}\a J \right] d_1
=0
\label{system}
\ena
First we observe that in the classical limit
$\a \to 0$ all the constants $G,H,I,J$ disappear and
the previous equations admit the classical
solutions for the boundary coefficients
\bea
&~& {\rm a)} ~~d_0 = d_2 = 0 ~~~~~~~~~~d_1~~{\rm {arbitrary}}
\nonumber \\
&~& {\rm b)} ~~d_1 = \pm \sqrt{2} ~~~~~~~~~~~~d_0,~d_2 ~~{\rm {arbitrary}}
\label{class}
\ena
in agreement with the results in Ref. \cite{corr}.

Then we study the system (\ref{system}) at the quantum level.
It is easy to verify that in the presence of a nonvanishing boundary
potential, neglecting
total derivative terms in the current ($G=H=I=J=0$)
would necessarily give unphysical (imaginary) solutions for
some of the boundary coefficients $d_j$.
Therefore acceptable solutions are obtained only for nontrivial
values of $G,H,I,J$. In particular it is possible to choose
these constants so that the equations in (\ref{system}) are satisfied by
the classical boundary coefficients in eq. (\ref{class}) with the extra
condition $d_0=d_2$. The details of the calculation and the explicit
solutions will be given in a future publication \cite{b6}.

\vspace{.8cm}
{\em{The $c^{(1)}_2$ affine Toda theory}}

\vspace{.2cm}
\noindent
The lagrangian for this nonsimply--laced theory is given by
\EQ
\b^2 {\cal L}= \theta(x_1) \left[ \frac12  \pa_{\mu} \phi_1 \pa_{\mu} \phi_1
+\frac12 \pa_{\mu}\phi_2 \pa_{\mu} \phi_2  + e^{-\sqrt{2}(\phi_1+\phi_2)}
+ 2e^{\sqrt{2}\phi_2} + e^{\sqrt{2}(\phi_1-\phi_2)}\right] -\d(x_1) B
\label{action2}
\EN
where the boundary perturbation $B$ is
\EQ
B=  d_0 ~e^{- \frac{1}{\sqrt{2}} (\phi_1+\phi_2 )}
+ d_1~ e^{\frac{1}{\sqrt{2}}\phi_2}
+d_2 ~ e^{\frac{1}{\sqrt{2}}(\phi_1-\phi_2)}
\label{pot2}
\EN
The first nontrivial high--spin conserved current in the bulk region
is at spin 4: the most general expression consistent with the symmetry of
the lagrangian under
$\phi_1\rightarrow -\phi_1$, is
\bea
J^{(4)}&=&A(\pa \phi_1)^4+B(\pa \phi_2)^4+C(\pa \phi_1)^2(\pa \phi_2)^2 +
+D\pa\phi_1\pa^2\phi_1 \pa \phi_2 +E(\pa^2 \phi_1)^2
+ F(\pa^2 \phi_2)^2 +\nonumber\\
{}~&~& +G\pa((\pa \phi_1)^2 \pa \phi_2)
+H\pa( \pa \phi_2)^3 + I\pa(\pa\phi_1\pa^2\phi_1)
+J\pa(\pa\phi_2\pa^2\phi_2)
\label{current2}
\ena
Cancellation of anomalous contributions to the conservation
equation in the bulk fixes the $A,B,\dots,F$ coefficients in (\ref{current2})
(see also \cite{b5})
\bea
A &=& B~=1~~~~~~~~~~~~~~~~C=-6(1+\a)~~~~~~~~~
{}~~~~~~~~~D=-6\sqrt{2}(2+3\a+\a^2)  \nonumber\\
E &=& -(4+12\a+\frac{23}{2}\a^2+3\a^3)~~~~~~~~~~~~~~~~~
F=2+3\a+\frac{\a^2}{2}
\ena
leaving $G, H,I,J$ undetermined. A lengthy calculation gives
the quantum trace (with the definition in (\ref{trace}))
\bea
\Theta_0&=&\left\{ \left[-(2+\frac{17}{3}\a+\frac{9}{2}\a^2+\frac{5}{6}\a^3)
+\frac{\a^2}{2\sqrt{2}} (G+H) -\frac{\a}{2} (I+J) \right]
(\pa\phi_1+\pa\phi_2)^2+ \right. \nonumber\\
{}~&~&~+\left[\frac{1}{\sqrt{2}}(1 + 2\a)G - I \right]
(\pa\phi_1)^2 +\nonumber\\
{}~&~&~+\left[ \frac{\a}{\sqrt{2}} G+
\frac{3}{\sqrt{2}}(1+\a)H-J \right] (\pa\phi_2)^2 +\nonumber\\
{}~&~&~+\left[\sqrt{2}(1+\frac{3}{2}\a)G
+\frac{3}{\sqrt{2}}\a H-I-J \right]
\pa\phi_1 \pa\phi_2 ~+\nonumber\\
{}~&~&~ \left. -\left[\frac{\a}{\sqrt{2}}(\frac{1}{3}
-\frac{\a}{2}-\frac{5}{6}\a^2)+\frac{\a^2}{4}(G+H)+\frac{\a}{2\sqrt{2}}(I+J)
\right] (\pa^2\phi_1+\pa^2\phi_2)+ \right. \nonumber \\
{}~&~&+ \left.
\frac{1}{\sqrt{2}}I~\pa^2 \phi_1 + \frac{1}{\sqrt{2}}J~ \pa^2 \phi_2 \right\}
e^{-\sqrt{2}(\phi_1+\phi_2)} \nonumber\\
&~&~~~~~~~~~~~~\nonumber \\
\Theta_1 &=&\left\{ \left[ 6(2+3\a+\a^2)-\sqrt{2}G
\right] \right. (\pa\phi_1)^2+ \nonumber\\
{}~&~&~\left. +\left[ -2(2+\frac{10}{3}\a+\frac{3}{2}\a^2
+\frac{\a^3}{6})-\sqrt{2}(3+3\a+\frac{\a^2}{2})H
-2(1+\frac{\a}{2})J\right] (\pa\phi_2)^2 + \right. \nonumber\\
{}~&~&~-\left.\left[ \frac{\sqrt{2}}{3}(\a+\frac{3}{2}\a^2
 +\frac{\a^3}{2})+\frac{\a^2}{2} H+\sqrt{2}
(1+\frac{\a}{2})J\right] \pa^2\phi_2\right\} e^{\sqrt{2}\phi_2}\nonumber \\
&~&~~~~~~~~~~~\nonumber \\
\Theta_2 &=&\Theta_0 (\phi_1 \rightarrow -\phi_1)
\label{trace2}
\ena
Again, imposing the boundary condition (\ref{39}) one obtains
that no anomalous contributions are produced
if the following equations are satisfied
\bea
&& \a (G-3H)d_0=0 \nonumber\\
&~&~~~~~~~~~~~~~~~~~~~~~~~ \nonumber \\
&&(6+\frac{3}{2}\a-\frac{9}{2}\a^2)d_1d_0^2
-[12+23\a+\frac{41}{3}\a^2+\frac{9}{2}\a^3+\frac{5}{6}\a^4+ \nonumber\\
&&~~~~~~~~~~~~~~~ -\sqrt{2}\a(3+3\a+\a^2)H+\frac{\a}{2}(1+\a)I+
\frac{\a^2}{2}J]d_1=0 \nonumber\\
&~&~~~~~~~~~~~~~~\nonumber \\
&&3\a^2d_1^2d_0-\a[16+\frac{80}{3}\a+6\a^2-\frac{2}{3}\a^3
-\sqrt{2}(6+6\a+\a^2)H-2(2+\a)I]d_0=0\nonumber\\
&~&~~~~~~~~~~~~~~~\nonumber \\
&&\a (G-3H)d_2=0 \\
&~&~~~~~~~~~~~~~~~\nonumber \\
&&(6+\frac{3}{2}\a-\frac{9}{2}\a^2)d_1 d_2^2
-[12+23\a+\frac{41}{3}\a^2+\frac{9}{2}\a^3+\frac{5}{6}\a^4\nonumber\\
&&~~~~~~~~~~ -\sqrt{2}\a(3+3\a+\a^2)H+\frac{\a}{2}(1+\a)I+\frac{\a^2}{2}J]
d_1=0 \nonumber\\
&~&~~~~~~~~~~~~~~~\nonumber \\
&&3\a^2d_1^2d_2-\a[16+\frac{80}{3}\a+6\a^2-\frac{2}{3}\a^3
-\sqrt{2}(6+6\a+\a^2)H-2(2+\a)I]d_2=0 \nonumber
\label{system2}
\ena
In the classical limit the constants $G,\dots, J$ do not
enter and the solutions for the boundary coefficients are (see \cite{corr})
\bea
&~& {\rm a)} ~~d_1 = 0 ~~~~~~~~~~~~~~~~~~~~~~d_0,~d_2 ~~{\rm {arbitrary}}
\nonumber \\
&~& {\rm b)} ~~d_0,~d_2 = \pm \sqrt{2} ~~~~~~~~~~d_1 ~~{\rm {arbitrary}}
\label{class2}
\ena
At the quantum level the situation is similar to the $d_3^{(2)}$ case.
If we set $G= \dots= J=0$ we obtain inconsistent
results, whereas with nonvanishing total
derivatives we can satisfy the equations (\ref{system2}) without
modifying the classical value of the boundary coefficients \cite{b6}.

\vspace{.4cm}

In conclusion we have found that for the two nonsimply--laced Toda theories
under consideration, the coefficients of the total derivative terms in the
spin--4 current can be chosen appropriately so that they
have a finite (but not zero) classical limit and in general they depend
on the particular value of the boundary coefficients in (\ref{class}) and
(\ref{class2}) respectively. We emphasize once again that it is the
presence of these total derivative terms that allows to maintain the
quantum conservation of the $q^{(3)}$ charge while keeping the boundary
perturbing potential as fixed by the classical conservation. We observe
that this was not possible for the spin--3 current of the $a_n^{(1)}$
theories:
in this case if we impose the quantum conservation, even
including total derivative terms, we are forced to modify the interaction
at the boundary by a finite renormalization \cite{b4,b6}.

\vspace{1.0cm}
\noindent
This work has been partially supported by grants no. SC1--CT92--0789 and no.
CEE--CHRX--CT92--0035.

\end{document}